\documentclass[aps,prd,twocolumn,floatfix,nofootinbib,showpacs,superscriptaddress]{revtex4-1}
\usepackage{amsmath,amsfonts,amssymb,bm}
\usepackage{graphicx}
\usepackage{subfigure}
\usepackage{color}
\definecolor{purple}{rgb}{0.5,0,0.5}
\definecolor{blue}{rgb}{0.0,0,0.9}
\usepackage[colorlinks=true, pdfstartview=FitV, linkcolor=purple, citecolor= purple, urlcolor=blue]{hyperref}

\begin{document}
\title{A Modification of Faddeev-Popov Approach free from Gribov Ambiguity}
 
  \author{Chong-yao Chen }
\affiliation{Department of Physics and State Key Laboratory of Nuclear Physics and Technology,
Peking University, Beijing 100871, China}

 \author{Fei Gao }
\affiliation{Department of Physics and State Key Laboratory of Nuclear Physics and Technology,
Peking University, Beijing 100871, China}
\affiliation{Collaborative Innovation Center of Quantum Matter, Beijing 100871, China}

\author{Yu-xin Liu }
\email[Corresponding author: ]{yxliu@pku.edu.cn}
\affiliation{Department of Physics and State Key Laboratory of Nuclear Physics and Technology,
Peking University, Beijing 100871, China}
\affiliation{Collaborative Innovation Center of Quantum Matter, Beijing 100871, China}
\affiliation{Center for High Energy Physics, Peking University, Beijing 100871, China}

\date{\today}

\begin{abstract}
We propose a modified version of the Faddeev-Popov quantization approach for non-Abelian gauge field theory to avoid the Gribov ambiguity.
We show that by means of introducing a new method to insert the correct identity into the Yang-Mills generating functional and considering the identity generated by an integral through a subgroup of the gauge group, the problem of the Gribov ambiguity can be removed naturally.
Meanwhile by handling the absolute value of the Faddeev-Popov determinant with the method introduced by Williams and collaborators,
we lift the Jacobian determinant together with the absolute value and obtain a local Lagrangian. The new Lagrangian have a nilpotent symmetry which can be viewed as an analogue of the BRST symmetry.
\end{abstract}

\maketitle





\medskip

\section{Introduction}

It is known that standard Faddeev-Popov(FP) gauge fixing approach to non-Abelian gauge theory suffers from the Gribov ambiguity, which is caused by the multiple solutions of the gauge fixing condition in the non-Abelian gauge transformation~\cite{gribov1978quantization}.
This Gribov ambiguity breaks the identity inserted by Faddeev and Popov and leads in turn the approach to be flawed.
There have been a lot of attempts  to solve this problem (for a pedagogical review, see Ref.~\cite{vandersickel2012gribov}).
The most successful way among them is the Gribov-Zwanziger approach~\cite{gribov1978quantization,zwanziger1989local,zwanziger1993renormalizability}.
They propose that the functional integral domain of the generating functional should be constrained in the so-called Gribov region where the FP determinant is positive definite.
Gribov first introduces the no-pole condition by a semi-classical way,
and gets a non-local action by making use of a Heaviside step function~\cite{gribov1978quantization}.
Then, Zwanziger proposes another technique by analyzing the lowest eigenvalue of the FP operator which should be setting positive inside the Gribov region.
Such a technique results in the horizon condition, and achieves a local renormalizable Lagrangian, referred commonly Gribov-Zwanziger (GZ) Lagrangian~\cite{zwanziger1989local,zwanziger1993renormalizability}. The equivalence between the no-pole condition and the horizon condition has been checked~\cite{gomez2010equivalence,capri2013allorder}. The renormalizability of the GZ action can be proved to all order by using algebraic renormalization~\cite{sobreiro2004gribov,dudal2010more}.

In the standard GZ approach, the GZ parameter leads to a soft breaking of the Becchi-Rouet-Stora-Tyutin (BRST) symmetry.
This soft breaking  allows one to prove that the new parameter occurs in the GZ action, i.e. Gribov parameter, is physical~\cite{dudal2008refinement},
and signifies the possibility to introduce non-perturbative effects in infrared region~\cite{baulieu2009soft},
which can be the reason that the standard GZ action brings in the enhanced behaviour of the ghost propagator in infrared region as the Kugo-Ojima confinement criterion~\cite{kugo1979covariant}, while it does not coincide with recent lattice QCD simulation results in large volume~\cite{bogolubsky2007lat,cucchieri2007lat,sternbeck2007lat},
that is,  a finite ghost in the infrared region.
This soft breaking is then identified as a spontaneous breaking of the BRST symmetry~\cite{dudal2012gribov,capri2013spontaneous}.
It has also been shown that the soft breaking part of the BRST symmetry can be compensated by adding a Gribov parameter dependent variation into the ordinary BRST transformation~\cite{dudal2011brst,capri2015exact,capri2016local}.
Moreover, after adding a quadratic term into the GZ action~\cite{dudal2008refinement},
the refined action leads the behaviour of the gluon and ghost propagators in infrared region to meet the latest lattice QCD result.
It has been shown that with the Dyson-Schwinger equation approach~\cite{aguilar2008gluon,boucaud2008on} or the functional renormalization group (FRG) approach~\cite{fischer2009on}, one can also get the same result.

Although the GZ approach has achieved a lot in latest decades, it is still not a complete solution for the Gribov ambiguity since the Gribov region still contains Gribov copies~\cite{vandersickel2012gribov,maskawa1978dense,semenov1986variational,vanbaal1992more}.
This problem is severe in lattice QCD simulations, since the numerical result will be logarithm dependent~\cite{cucchieri1997gribov}.
A further constraint to the so-called fundamental modular region (FMR, also called minimal Landau gauge or absolute Landau gauge in lattice QCD~\cite{giusti2001problems}) seems necessary, but there is still no analytic expression for it~\cite{vandersickel2012gribov}.
Another difficulty in lattice QCD simulation is the enormous consumption of calculation resource, to simplify
the numerical calculation,
the maximal Abelian gauge  was proposed (e.g., Ref.~\cite{shinohara2003most}).
However, it also suffers from the problem of Gribov ambiguity~\cite{capri2015gribov}.
Alternatively, there are some works trying to analyze the different contribution of Gribov copies~\cite{serreau2014covariant}, or averaging all Gribov copies~\cite{smekal2007modified,smekal2008lattice}.
Generally speaking, the Gribov region always makes the contribution of Gribov copies more unmanageable.

Besides constraining configuration space to some subset, there are also other ways,
for instance, finding a gauge fixing condition which is free from Gribov copies, such as space-like planar gauge~\cite{bassetto1983absence}, modified axial gauge~\cite{zhou2017quantization}, and so on. However, such gauge conditions either break the Lorentz invariance  or violate the infinite vanishing boundary condition~\cite{singer1978some,vandersickel2012gribov}.
The Gribov-copy-free, continuous gauge fixing condition which satisfies the infinite vanishing boundary condition is still imperatively expected.
Or one can consider adding lorentz covariant effective potential terms of a large number of auxiliary fields to the Yang-Mills action to avoid the Gribov ambiguity~\cite{slavnov2008lorentz,slavnov2012study}, but these additional fields make the theory more complicated.

In this paper we will return to the original Faddeev-Popov approach.
We show that by reconsidering the true identity that should be inserted into the Yang-Mills (YM) generating functional, one can get a local action which is free from the Gribov ambiguity. By implementing the method that handles the absolute value sign of the Faddeev-Popov determinant introduced by Ghiotti and collaborators in Ref.~\cite{ghiotti2005landau}, we get a localized Lagrangian and show that the new Lagrangian holds a nilponent symmetry which is an analogue of the BRST symmetry.

\section{Gauge Symmetry of Yang-Mills Theory}

At first we elucidate some conventions and definitions, which might be slightly different from the usual case.

Throughout this paper, we focus on the pure Yang-Mills theory based on a given semi-simple compact Lie group $G$, without taking matter fields into account. The action of pure $G$-Yang-Mills theory is
\begin{equation}\label{lag}
S_{\mathrm{YM}}=\frac{1}{2g^{2}}\int\mathrm{d}^{d}xTr(F^{\mu\nu}F_{\mu\nu})
=\frac{1}{4g^{2}}\int\mathrm{d}^{d}xF^{\mu\nu,a}F^{a}_{\mu\nu},
\end{equation}
were $F_{\mu\nu}=\partial_{\mu}A_{\nu}-\partial_{\nu}A_{\mu}+[A_{\mu},A_{\nu}]$ is the field strength tensor with $A_{\mu}$ the gauge field,  which can be decomposed as: $A_{\mu}=igA^{a}_{\mu}t^{a}$
with $t^{a}$  the infinitesimal generators of the group $G$.
The infinitesimal generators connects to the base of the corresponding Lie algebra $\mathfrak{g}$
with relation $e^{a}=igt^{a}$, $A_{\mu}$ is thus a $\mathfrak{g}$-value function.
We denote the set of all the $A$ as $\mathfrak{A}$, called configuration space,
where rigorously speaking $A=A_{\mu}\otimes\mathrm{d}x^{\mu}$ is a $\mathfrak{g}$-value 1-form. We can also simply consider it as the ordered set of $A_{\mu}$. There are some other constraints for $A_{\mu}$, i.e. the infinity vanishing boundary condition, which insures that there is no Gribov copies encounter in QED ($U(1)$-Yang-Mills theory). Therefore, $\mathfrak{A}$ is a subspace of the vector space containing all the $\mathfrak{g}$-value 1-form on some $d-$dimensional time-space manifold $M$~\cite{atiyah1978self}.

The YM action is invariant under local gauge transformation
\begin{equation}\label{gt}
A_{\mu}^{} \longrightarrow A_{\mu}^{U}=UA_{\mu}U^{\dagger}+(\partial_{\mu} U)U^{\dagger},
\end{equation}
where $U(x)$ is an element of the group $G$ for any space-time coordinate $x$.
In other word, it is a $G$-value function.
It also forms a group owing to the properties of the pointwise group multiplication and inverse: $U_{1}(x)\cdot U_{2}(x)=U_{1}\cdot U_{2}(x);(U(x))^{-1}=U^{-1}(x)$.
We denote this group as: $\mathcal{G}$, the gauge group.
We can also write $U(x)$ as: $e^{X}$,
here $X=ig\theta^{a}t^{a}=\theta^{a}e^{a}$ is a $\mathfrak{g}$-value function.
By the virtue of the relation between $G$ and $\mathfrak{g}$ we get
\begin{equation}\label{gtn}
\begin{split}
A^{U}_{\mu}&=e^{\mathrm{ad}_{X}}A_{\mu}+(\partial_{\mu}e^{X})e^{-X}\\
&=\frac{e^{\mathrm{ad}_{X}}-1}{\mathrm{ad}_{X}}(\partial_{\mu}X+\mathrm{ad}_{X}A_{\mu})+A_{\mu}\\
&=A_{\mu}+\frac{e^{\mathrm{ad}_{X}}-1}{\mathrm{ad}_{X}}(D_{\mu}X),
\end{split}
\end{equation}
where we have already used the identities $e^{\mathrm{ad}_{X}}(Y)=e^{X}Ye^{-X}$, and $\partial_{\mu}e^{X}=e^{X}(\frac{1-e^{\mathrm{-ad}_{X}}}{\mathrm{ad}_{X}}(\partial_{\mu}X))$. We can also rewrite the covariant derivation as $D_{\mu}=\partial_{\mu}-\mathrm{ad}_{A_{\mu}}$ which is useful in the following.

If we take $\theta^{a}$ infinitesimal, Eq.~(\ref{gtn}) can be simplified as
\begin{equation}
A^{U}_{\mu}=A_{\mu}+D_{\mu}X.
\end{equation}
In adjoint representation, the covariant derivation is $D^{ab}_{\mu}=\partial_{\mu}\delta^{ab}+f^{abc}A_{\mu}^{c}$,
with  $f^{abc}$ the structure constant of the group $G$.

Furthermore, we introduce the definition of gauge orbit, i.e. gauge equivalent class.
Configurations $A$ and $A^{\prime}$ are said to be gauge equivalent if $A_{\mu}^{U}=A^{\prime}_{\mu}$, we denote it as $A^{\prime} \sim A$, which stands for an equivalent relation. All the configurations equivalent to $A$ form a set $[A]$, called the gauge orbit.
If a quantity $Q$ depends only on the orbit but not the specific configuration, i.e. gauge invariant quantity,
we denote it as $Q[A]$. For more information about the proposition of the gauge orbit one can refer to Ref.~\cite{rudolph2017differential}.

\section{Overview of Gribov Problem and Gribov-Zwanziger Approach}

For the convenience of detailed discussion, we overview the Gribov problem which occurs in the procedure of quantizing the Yang-Mills action, which is based on Ref.~\cite{vandersickel2012gribov}.
People have found that in the standard path integral quantization of gauge field theory, the non-interacting propagator of gauge field is ill-defined, which arises from the fact that the gauge equivalent configurations are over-counted in the gauge invariant functional integral, therefore the gauge should be fixed.
Faddeev and Popov proposed then the approach of inserting an identity
\begin{equation}\label{id21}
1=\int\mathcal{D}U\delta(G(A^{U}))\det\frac{\delta(G^{a}(A^{U,\theta}))}{\delta\theta^{b}}\Big{\vert}_{\theta=0}
\end{equation}
into the YM generating functional $Z=\int\mathcal{D}Ae^{\alpha S_{YM}}$, where $G(A)$ is the gauge fixing condition, usually is a set of differential equations, $A^{U,\theta}$ represents applying an infinitesimal gauge transformation, parameterized by $\theta$, to the gauge transformed $A^{U}$. The parameter $\alpha$ on the exponent is varying along with the different choice of metric signature, e.g. $\alpha = i$ for Lorentzian, $\alpha = -1$ for Eulidean. For simplicity, we will working in $d$-dimensional Minkowski space from now on.
The generating functional becomes then:
\begin{equation}\label{tra}
Z=\int\mathcal{D}A\int\mathcal{D}U\delta(G(A^{U}))\det\frac{\delta(G^{a}(A^{U,\theta}))}{\delta\theta^{b}}
\Big{\vert}_{\theta=0}e^{iS_{YM}}.
\end{equation}
By the virtue of gauge invariance property of $\mathcal{D}A$ and YM action, people can replace the variables in Eq.~(\ref{tra}) as $A^{U}\rightarrow A$, and extract the volume of gauge group which is an infinity constant. The generating functional can then be rewritten as:
\begin{equation}
Z=\int\mathcal{D}A\delta(G(A))\det\frac{\delta(G^{a}(A^\theta))}{\delta\theta^{b}}
\Big{\vert}_{\theta=0}e^{iS_{YM}}.
\end{equation}
In linear covariant gauge one can introduce ghost fields $c^{a},\bar{c}^{a}$ to deal with the functional determinant, and take the Gaussian distribution to treat the $\delta(G(A))$.
People get then the FP Lagrangian:
\begin{equation}
\mathcal{L}=\mathcal{L}_{YM}-\frac{1}{2\xi}(\partial^{\mu}A^{a}_{\mu})^{2}-\bar{c}^{a}(\partial^{\mu}D_{\mu}^{ab})c^{b}.
\end{equation}
However there are several problems in the FP approach, which is first referred by Gribov~\cite{gribov1978quantization}.

Gribov notices that the identity in Eq.~(\ref{id21}) is not always validating~\cite{gribov1978quantization}, and the exact identity should contains an absolute value sign which is due to the positive definiteness of delta function, and can be written in two different ways
\begin{equation}\label{coi}
1=\frac{1}{\sum_{U^{\prime},G(A^{U^{\prime}})
=0}\frac{1}{|\Delta(G(A^{U^{\prime}}))|}}\int\mathcal{D}U\delta(G(A^{U})),
\end{equation}
or, using $N(G[A])$ to denote the number of solutions for the differential equation: $G(A^{U})=0$, i.e. $N(G[A])=|S(G[A])|=|\{U\in\mathcal{G}:G(A^{U})=0\}|$
\begin{equation}\label{coi1}
1=\frac{1}{N(G[A])}\int\mathcal{D}U\delta(G(A^{U}))|\Delta(G(A^{U}))|,
\end{equation}
where $\Delta(G(A))=\det M^{ab}=\det\frac{\delta(G^{a}(A^{\theta}))}{\delta\theta^{b}}\mid_{\theta=0}$ is the determinant of the Faddeev-Popov operator, called FP determinant, with $a$ and $b$ the indices of group parameters for $U$.
The difference between the above two identities is that the FP determinant in Eq~(\ref{coi}) is evaluated at the fixed points, thus constant, but the counterpart in Eq.~(\ref{coi1}) is a functional along with $U$. Therefore if we take the second identity, as we will do, we need to keep in mind that the FP determinant is not a constant, but varying along with $U$.

To remedy the shortage in the FP approach, Gribov introduces the conception of Gribov copy:
the $A^{\prime}$ and $A$ are a pair of Gribov copies, if the two configurations are in the same gauge orbit $[A]$, and both satisfy the gauge fixing condition $G(A)=0$, and certainly, $A^{\prime}\neq A$.
One thing we need to emphasize is that the $N(G[A])$ can not be considered as the number of Gribov copies in the gauge orbit $[A]$, since $S(G[A])$ includes all the solutions that satisfy $A^{U}=A$. If we denote the number of Gribov copies for the orbit $[A]$ as $N_{GC}(G[A])$, there is a simple relation between $N_{GC}(G[A])$ and $N(G[A])$ by applying the orbit-stabilizer theorem (see, e.g. Ref.~\cite{hungerford1973algebra})
\begin{equation}
N(G[A])=|Z(A)|\cdot N_{GC}(G[A])=\frac{|\mathcal{G}|}{|[A]|}\cdot N_{GC}(G[A]),
\end{equation}
where $Z(A)$ is the stabilizer subgroup for configuration $A$, i.e. $Z(A)=\{U\in\mathcal{G}:A^{U}=A\}$, and $|Z(A)|=\frac{|\mathcal{G}|}{|[A]|}$  is a gauge invariant quantity according to the  orbit-stabilizer theorem. Therefore,  there exists difference between the $N(G[A])$ and the $N_{GC}(G[A])$ up to a orbit-dependent factor $|Z(A)|$. It means that the number of Gribov copies is not necessary to be identical for each orbit, and we can not extract $\frac{1}{N(G[A])}$ out of the functional integral.
As a consequence, people should use Eq.~(\ref{coi1}) as the new identity in FP approach, and the generating functional should be rewritten as
\begin{equation}\label{coi2}
Z=\int\mathcal{D}A\frac{1}{N(G[A])}\delta(G(A))|\Delta(G(A))|e^{iS_{\mathrm{YM}}}.
\end{equation}

From the above description one can see that the standard FP approach needs to be fixed from two aspects: one is that the absolute value should not be simply neglected,
the other is that the effect of the Gribov copies need to be analyzed.

Gribov introduces the Gribov region to fix these two problems, which is now called Gribov-Zwanziger approach. In Gribov's famous seminal article~\cite{gribov1978quantization}, he proposed that people can constrain the functional integral domain to Gribov region $\Omega$, where the FP determinant inside is positive definite, so that the first problem brought by the Gribov ambiguity can be removed.
The rigorous definition for the Gribov region reads then
\begin{equation}
\Omega=\{A_{\mu}:\partial_{\mu}A_{\mu}^{a}=0,M^{ab}>0\}.
\end{equation}
People can interpret the generating functional as
\begin{equation}\label{gr}
Z=\int\mathcal{D}AV(A)\delta(\partial^{\mu}A_{\mu})\Delta(G(A))e^{iS_{\mathrm{YM}}(A)},
\end{equation}
where $V(A)$ is a factor that represents the positive definite information of FP operator,
which Gribov interprets as a Heaviside step function.

However, there is still an unsolved problem in Gribov's procedure, that is, Gribov region still contains Gribov copies~\cite{maskawa1978dense,semenov1986variational},
which makes the relation between the new and the original generating functional even more complicated. Therefore, in this paper, we will not follow the route of the GZ approach.

Besides, there is another issue easily to be neglected. It occurs when people try to lift the $\delta(G(A))$ to the exponential factor. To see it through we will work in linear covariant gauge $\partial^{\mu}A^{a}_{\mu}=\omega^{a}$. By implementing an integral through Gaussian distribution of $\omega$, people get
\begin{equation}
Z=\int\mathcal{D}A\mathcal{D}\omega e^{-\frac{\omega^{2}}{2\xi}}\frac{1}{N(G[A])}\delta(\partial^{\mu}A^{a}_{\mu}-\omega^{a})|\Delta(G(A^{U}))|e^{iS_{\mathrm{YM}}},
\end{equation}
where the easily ignorable thing is, $N(G[A])$ depends on $\omega$ although the FP determinant is independent from $\omega$, which means that the factor $N(G[A])=|\{U\in\mathcal{G}:\partial^{\mu}A^{U}_{\mu}=\omega\}|$ is not only gauge orbit dependent, but also $\omega$-dependent. Generally speaking, this $\omega$-dependence cannot be expressed as a localized term in the Lagrangian, only if we let $\xi \rightarrow 0$. After then, the divergence of $A_{\mu}$ is fixed, and thus $N([A],\partial^{\mu}A^{a}_{\mu})$ depends only on the gauge orbit, which is much simpler. This means that when the the problem of Gribov ambiguity occurs, the localized covariant Lagrangian can only be obtained in Landau gauge $\partial^{\mu}A^{a}_{\mu}=0$.
If one considers only to solve the problem caused by the absolute value (e.g. GZ approach), it has been shown possible to work in general $R_{\xi}$ gauge~\cite{lavrov2011soft,lavrov2013gribov}. \\

After having reviewed the problems brought by Gribov ambiguity explicitly,
we will find a way to avoid them now.

\section{A New Approach to Quantize the YM Field}

Let us recall the original idea of FP approach. The key point is to insert an identity that equals to one, and then extract the infinity related to the integral of the gauge parameters.  However, the Gribov copies make the identity invalid. The direct way to eliminate the Gribov copies is constraining the configuration inside the Gribov region, but there are many obstacles to proceed such an approach in a correct way. Here we propose a new approach complying with the original FP method, that is, to insert the true identity which holds even when there are Gribov copies.

It is evident that the main idea of the identity in Eq.~(\ref{coi1}) is that the functional integral domain does not need to be the complete gauge group, if we change the domain, this identity will still hold.
Therefore, we will now apply a smaller group for this integral, as long as the infinity factor can still be extracted.
Noticing that after inserting this identity, we need to change the integral variable by: $A^{U}_{\mu}=A_{\mu}$, thus the new domain we choose is better to be a subgroup of the gauge group $\mathcal{G}$.
A natural choice is the subgroup $\mathcal{H}$ pointwisely generated by the Cartan subalgebra $\mathfrak{h}$ of $\mathfrak{g}$.
In other word, for any time-space point $x$, $U(x)$ now belongs to the Abelian subgroup of $G$, generated by the pairwisely commutative generators.
We denote the Lie subalgebra corresponding to the $\mathcal{H}$ as $\mathcal{X}$.
Although throughout this article, we refer to $\mathcal{H}$ as the maximal Abelian subgroup, such an $\mathcal{H}$ may in fact not be the maximal Abelian subgroup of $\mathcal{G}$.
Detailed discussion about the case that $\mathcal{H}$ is definitely not the maximal Abelian subgroup of $\mathcal{G}$ will be given elsewhere.

Now, the whole generating functional is formally the same, but for mathematical validity the term $\delta(G(A^{U}))$ needs some modification and also the FP determinant $|\Delta(G(A^{U}))|$. The generating functional can be rewritten as
\begin{equation}
Z=\int\mathcal{D}A\frac{1}{N(G[A])}\int_{\mathcal{H}}\mathcal{D}U\delta(G(A^{U}))|\Delta(G(A^{U}))| e^{iS_{\mathrm{YM}}}.
\end{equation}
Here the meaning of $N(G[A])$ is also changed, so as $N(G[A])=|\{A^{U}:G(A^{U})=0,U\in\mathcal{H}\}|$.
Next, we follow the standard FP approach, changing the configuration integral variable through gauge transformation: $A^{U}_{\mu}=A_\mu$, and thus extracting the infinity constant $\int_{\mathcal{H}}\mathcal{D}U$. The generating functional becomes:
\begin{equation}
Z=\int\mathcal{D}A\frac{1}{N(G[A])}\delta(G(A))|\Delta(G(A))|e^{iS_{\mathrm{YM}}}.
\end{equation}\par
From now on, we choose the gauge group to be $SU(2)$ as an example for detailed discussion.
The Abelian subgroup $\mathcal{H}$ now is simply $U(1)$, which is generated by a single parameter $\theta$.
Since the dimension of the term $\delta(G(A))$ should be identical to the dimension of the group integral domain, we change our gauge condition $G(A)$ to be quadratic gauge $(\partial^{\mu}A^{a}_{\mu})^{2}=\omega$, which is equivalent to the ordinary Landau gauge,
their solutions consist with a hyperplane in configuration space $\mathcal{A}$, call Landau hypersurface $R$. The generating functional is now
\begin{equation}
Z=\int\mathcal{D}A\frac{1}{N(G[A])}\delta((\partial_{\mu}A^{a\mu})^{2}-\omega) |\Delta(G(A))|e^{iS_{\mathrm{YM}}}.
\end{equation}
Since the generating functional is independent of the choice of gauge fixing condition, we can introduce any distribution. In present case we choose it as $e^{-\frac{i\omega}{2\xi}}$, and integrate over $\omega$, we have
\begin{eqnarray}
\nonumber
Z = \!\! & \!\! \int\mathcal{D}A\frac{1}{N(G_{\omega}[A])}\int\mathcal{D}\omega e^{-\frac{i\omega}{2\xi}}\delta((\partial_{\mu}A^{a\mu})^{2}-\omega)\\[1mm]  \nonumber
&  \qquad \times |\Delta(G(A^{}))|e^{iS_{\mathrm{YM}}}\\   
 = \!\! & \!\! \int\mathcal{D}A \frac{1}{N(G_{(\partial_{\mu}A^{a\mu})^{2}}[A])}|\Delta(G(A^{}))|
e^{iS_{\mathrm{YM}}-i\frac{(\partial^{\mu}A^{a}_{\mu})^{2}}{2\xi}}.
\end{eqnarray}
As the same as we have discussed in the last section, only when $\xi\rightarrow0$ can we get a certain value for $N(G_{\omega}[A])$, which is obviously equivalent to Landau gauge fixing condition $\partial^{\mu}A_{\mu}=0$. In this circumstance we can prove that the $N(G[A])=1$.

We choose the base $t^{a},a=1,2,3$ of $su(2)$ as usual, where $t^3$ is the Abelian generator,
we have thus the commutative relations
\begin{equation}\label{com}
[t^{3},t^{1}]=t^{2}, \quad [t^{3},t^{2}]=-t^{1}.
\end{equation}

Recall the local gauge transformation in Eq.~(\ref{gtn}), and insert it into the equation
\begin{equation}\label{func}
\partial^{\mu}A_{\mu}^{U}=\partial^{\mu}A_{\mu}=0,
\end{equation}
we have
\begin{equation}
\partial^{\mu}\frac{e^{\mathrm{ad}_{X}}-1}{\mathrm{ad}_{X}}(\partial_{\mu}X+[X,A_{\mu}])=0.
\end{equation}
Now, we need to solve this equation, for $X=\theta t^{3}$, since $[X,\partial_{\mu}X]=0$. Expanding $A_{\mu}=A^{a}_{\mu}t^{a}$ we have
\begin{equation}
\partial^{2}\theta t^{3}+\partial^{\mu}(A^{b}_{\mu}(e^{\mathrm{ad}_{X}}-1)(t^{b}))=0, \quad b=1,2 .
\end{equation}
The above equation can be decomposed into three linear independent equations.
For $t^{3}$ we have
\begin{equation}
\partial^{2}\theta=0 ,
\end{equation}
where we have used $(e^{\mathrm{ad}_{X}}-1)(t^{3})=0$ and $(e^{\mathrm{ad}_{X}}-1)(t^{1,2})$ is irrelevant from $t^{3}$ due to the commutative relations in Eq.~(\ref{com}).
Thus $\theta$  have only  trivial global solutions and plane-wave solution, which is forbidden by the infinity vanishing boundary condition.
Therefore, in our new approach Landau gauge is no longer suffering from the Gribov ambiguity in this smaller group. An issue needs to be mentioned is that if the configuration $A$ we choose in Eq.~(\ref{func}) is not located on the Landau hypersurface, there is a chance that the equation  $\partial^{\mu}A^{U}_{\mu}=0$ does not have any solution, so $N(G[A])$ may equals to 0.
We will see in section V that this can be avoided by a mathematical technique.

Now the identity we will insert can be written as
\begin{equation}\label{nid}
1=\int_\mathcal{H}\mathcal{D}U\delta(G(A^{U}))|\Delta(G(A^{U}))|.
\end{equation}
In this way, after having neglected the infinite constant volume of $\mathcal{H}$, the whole generating functional reads
\begin{equation}
Z=\int\mathcal{D}A|\Delta(G(A))|e^{i \big{(} S_{\mathrm{YM}}-\frac{(\partial^{\mu}A^{a}_{\mu})^{2}}{2\xi} \big{)}}.
\end{equation}
This expression is very similar to the one in the original FP approach, except that here it is the subgroup transformation and also there is an additional absolute.
Next, we need to deal with the determinant $|\Delta(G(A))|$  with
\begin{equation}
\Delta(G(A))=\frac{\delta((\partial_{\mu}A^{\mu,a,\theta})^{2})}{\delta\theta} \Big{\vert}_{\theta=0} .
\end{equation}

Noticing that the infinitesimal version of a local gauge transformation is
$A^{a\theta}_{\mu}=A^{a}_{\mu}+D_{\mu}^{a3}\theta$, $D^{a3}_{\mu}=\partial_{\mu}\delta^{a3}+\epsilon^{ab3}A_{\mu}^{b}$, we get
\begin{equation}\label{det28}
\Delta(G(A))=\det[(\partial_{\mu}A^{a\mu})(\partial^{\mu}D^{a3}_{\mu})]=\det M,
\end{equation}
where $\epsilon^{abc}$ is the structure constant of the group $SU(2)$ in adjoint representation to distinguish that, namely $f^{abc}$, in the general case with gauge group $\mathcal{G}$. The term $\partial_{\mu}A^{a\mu}$ tends to zero, but yet since it marks the difference between the contributions from the $a=1,2,3$, it is necessary to preserve it. This functional determinant can be dealt with Grassmann fields, $c,\bar{c}$ (the ghost field):
\begin{equation}\label{det29}
\Delta(G(A))=\int\mathcal{D}\bar{c}\mathcal{D}c \exp\Big{\{}-\int d^{d}x \bar{c}(\partial_{\mu}A^{a\mu})
(\partial^{\mu}D^{a3}_{\mu})c \Big{\}} .
\end{equation}
The introduced ghost field holds zeroth mass scale and no quadratic kinematic term.
We need to remember there is absolute value sign of $\Delta(G(A))$ that contributes to the generating functional,
and although it is approaching zero when the gauge fixing condition is satisfied,
the whole integrand is finite, and the absolute value sign cannot be taken out.
One can easily understand it by considering the analogue integral in real number: $\int\mathrm{d}x\delta(x^{2})|2x|=\int\mathrm{d}x\delta(x)=1$.
In the following, to deal with the absolute value sign, we follow the method introduced in Ref.~\cite{ghiotti2005landau}
\begin{equation}
|\Delta(G(A))|=\mathrm{sgn}(\Delta(G(A)))\Delta(G(A)).
\end{equation}
Since we have had the $\Delta(G(A))$ as Eq.(\ref{det29}), what we need to do is just to find an expression for $\mathrm{sgn}(\Delta(G(A)))$. By considering the Lagrangian
\begin{equation}
\mathcal{L}_{\mathrm{sgn}}=BM\varphi-\bar{d}Md+\frac{1}{2}B^{2},
\end{equation}
where $M= (\partial_{\mu}A^{a\mu})(\partial^{\mu}D^{a3}_{\mu})$, $\bar{d}$ and $d$ are new Grassmannian fields, $\varphi$ and $B$ are auxiliary commuting Hermitian fields, we have:
\begin{equation}
\begin{split}
Z_{\mathrm{sgn}}&=\int\mathcal{D}\bar{d}\mathcal{D}d\mathcal{D}\varphi\mathcal{D}Be^{i\int d^{4}x\mathcal{L}_{\mathrm{sgm}}}\\
&=\frac{\det M}{\sqrt{\det(M^{\mathrm{T}}M)}}=\mathrm{sgn}(\det(M)).
\end{split}
\end{equation}
The full Lagrangian (which only stands for Landau gauge $\xi=0$) under our new approach reads now
\begin{equation}\label{la}
\mathcal{L}_{f}=\mathcal{L}_{\mathrm{YM}} -\frac{(\partial^{\mu}A^{a}_{\mu})^{2}}{2\xi}-\bar{c}Mc+BM\varphi-\bar{d}Md+\frac{1}{2}B^{2},
\end{equation}
with $M=(\partial_{\mu}A^{a\mu})(\partial^{\mu}D^{a3}_{\mu})$.
And the free propagator of the gauge field is well-defined as
\begin{equation}
\begin{split}
\langle A_{\mu}^{a}(x)A_{\nu}^{b}(y) \rangle = \int\frac{\mathrm{d}^{4}k}{(2\pi)^{4}}&\frac{-i}{k^{2}+i\epsilon}
\Big{(} g_{\mu\nu}-\frac{k_{\mu}k_{\nu}}{k^{2}} \Big{)}   \\
& \times\delta^{ab}e^{-ik\cdot(x-y)}  ,
\end{split}
\end{equation}
where we choose the time-space manifold as 4-dimensional Minkowski space.
From the above analysis working in the case of $SU(2)$ gauge symmetry specifically, we have shown apparently that, if one works in the maximal Abelian subgroup of the gauge group $\mathcal{G}$, there is no multiple solutions for the gauge fixing condition, i.e. no Gribov ambiguity occurs, as long as we insist on the infinity vanishing boundary condition. And the consequent bare gluon propagator is exactly the same as the counterpart working in the full gauge group $\mathcal{G}$.

This approach can be generalized into $SU(N)$ gauge theory, especially $SU(3)$,
the QCD case where the maximal Abelian subgroup is $U(1)\otimes U(1)$.
Now there are two Abelian generators, we need thus to choose a 2-dimensional gauge fixing condition
which should better be equivalent to Landau gauge.
The choice is not unique. We can arbitrarily distribute the color indices into two sets $\Lambda^{1,2}$, and let the gauge fixing condition be $\sum_{a\in\Lambda^{1}}(\partial^{\mu}A^{a}_{\mu})^{2}=\omega_{1},\sum_{b\in\Lambda^{2}}(\partial^{\mu}A^{b}_{\mu})^{2}=\omega_{2}$. %
It is worth to mention that it is permitted to put all the color indices into one single set, in such a case the maximal Abelian group $U(1)\otimes U(1)$ will degenerate into $U(1)$, just as that in the $SU(2)$ case.
Whatever the choice is, when $\omega_{1,2}\rightarrow0$ it tend to the Landau gauge.
Thus we can still take the same measure as used in the $SU(2)$ case to prove that this approach is free from Gribov ambiguity.
We can easily see further that the free propagator for the gauge field will be the same,
but the Jacobian part might change by the different choices of $\Lambda^{1,2}$. On the other hand, if we keep it in the subgroups $(U(1))^N$, it will be in the same expression for any $SU(N)$ group.

\section{New Nilpotent Symmetry and Some Remarks}

The new Lagrangian with gauge fixing is not gauge invariant anymore, which will lead to the similar situation in the standard Faddeev-Popov approach where introduces a nilpotent symmetry, i.e. BRST symmetry. It is natural to ask whether our new Lagrangian still preserves this symmetry in any sense.
The answer is yes. Following the procedure in Ref.~\cite{ghiotti2005landau}, we can get an new extend Nilpotent symmetry for the Lagrangian in Eq.~(\ref{la}), which is a little different from the usual meaning of a symmetry. For simplicity in the following discussion we take the condensed De-Witt notation~\cite{dewitt1965dynamical}.
Recalling the BRST transformation for $SU(N)-$gauge theory
\begin{equation}\label{brsttrans}
\Phi\rightarrow\Phi+\lambda s\Phi, \;\;\Phi\in\{A_{\mu}^{a},c^{b},\bar{c}^{c},b^{d}\},\;\;a,b,c,d=1,\cdots,N
\end{equation}
where $\lambda$ is an arbitrary constant Grassmanian coefficient (not necessarily being infinitesimal), and $s$ stands for the Slavnov operator which is the generator of the above transformation, more generally it can be understood as a linear operator defined on the polynomial algebra generated by $\{A_{\mu}^{a},c^{b},\bar{c}^{c},b^{d}\}$ and their all order differentials.
Besides, $s$ also abide by the Leibnitz law and commute with the differential operator.
The operation of $s$ acting on the zero-order generators can be expressed explicitly as:
$$   
sA_{\mu}^{a}=D_{\mu}^{ab}c^{b}, \;\; sc^{a}=-\frac{1}{2}gf^{abc}c^{b}c^{c}, \; \; s\bar{c}^{a}=b^{a}, \;\; sb^{a}=0, \;\;
$$
%
where the Lagrangian is written as
%
$$  \mathcal{L}_{\mathrm{FP}}=\mathcal{L}_{\mathrm{YM}}+b^{a}\partial^{\mu}A_{\mu}^{a}+\frac{\xi}{2}b^{a}b^{a} -\bar{c}^{a}\partial^{\mu}D_{\mu}^{ab}c^{b} .   $$
%
Though we now only have one of each ghost and anti-ghost field, more auxiliary fields encounter, we need thus a modification for the symmetry.
It is not hard to find that the first three terms of our new Lagrangian in Eq.~(\ref{la}) are invariant under transformation $\Phi\rightarrow \Phi+\lambda s^{\prime}\Phi $, with $s^{\prime}$ being defined as
\begin{equation}
\begin{split}
&s^{\prime} A_{\mu}^{a}=D_{\mu}^{a3}c, \;\; s^{\prime}c=0, \;\;\; s^{\prime}\bar{c}=-\frac{1}{\xi}, \\
&s^{\prime}\varphi=0,  \qquad \; s^{\prime}d=0, \;\;\; s^{\prime}\bar{d}=0, \;\;\; s^{\prime}B=0. \quad
\end{split}
\end{equation}
The last three terms in Eq.~(\ref{la}) are invariant under transformation $\Phi\rightarrow \Phi+\lambda t\Phi$, with $t$ defined as
\begin{equation}\label{2}
\begin{split}
&t A_{\mu}^{a}=0, \;\;\; tc=0, \;\;\;\; t\bar{c}=0, \\
&t\varphi=d,  \;\;\;\; td=0, \;\;\;\; t\bar{d}=B, \;\;\;\; tB=0. \quad
\end{split}
\end{equation}
It is apparent that the operator $t$ is nilpotent,
and to verify the nilpotency of the $s^{\prime}$, we should only prove $s^{\prime2} A_{\mu}^{a3} = 0$.
With the definition of the $s^{\prime}$, we have evidently
\begin{equation}
\begin{split}
{s^{\prime}}^{2} A_{\mu}^{a3} & = s^{\prime}D_{\mu}^{a3}c \\ &=\delta^{a3}\partial_{\mu}s^{\prime}c+s^{\prime}(\varepsilon^{a3b}A_{\mu}^{b}c)
=\varepsilon^{a3b} s^{\prime} (A_{\mu}^{b}) c  \qquad \\
&=\varepsilon^{a3b} \delta^{b3}\partial_{\mu}^{} c + \varepsilon^{a3b} \varepsilon^{b3c}A^{c}_{\mu} c\cdot c \, .
\end{split}
\end{equation}
Since $\varepsilon^{ab3}\delta^{b3} \equiv 0 $ by their definition and
$c\cdot c \equiv 0 $ due to its Grassmanian essence, ${s^{\prime}}^{2} A_{\mu}^{a3}$ must vanish,
i.e., the operator $s^{\prime}$ holds the nilpotency.

Now considering the diagonalized operator $\mathcal{S}=\mathrm{diag}(s^{\prime},t)$, which is of course also nilpotent, and rewriting Eq.~(\ref{la}) as
\begin{equation}
\mathcal{L}_{f}=\mathrm{Tr}\left( \!\!
\begin{array}{cc}
\mathcal{L}_{\mathrm{YM}} \! - \! \frac{(\partial^{\mu}A^{a}_{\mu})^{2}}{2\xi} \! - \! \bar{c}Mc   &  0 \\[1mm]
0 &  BM\varphi \! - \! \bar{d}Md \! + \! \frac{1}{2}B^{2} \\
\end{array}
\!\! \right)=\mathrm{Tr}L,
\end{equation}
the symmetry of Lagrangian Eq.~(33) is in the sense that $L$ remains unchanged under the transformation $L\rightarrow L+\lambda \mathcal{S}L$ due to $\mathcal{S}L=0$. Noticing the definitions, the nilpotency
and $tM = Mt$, one can prove such a symmetry directly.

In addition, noticing that in Landau gauge $\xi \rightarrow 0$, the transformation about $\bar{c}$ is ill-defined,
we then introduce the Nakanishi-Lautrup auxiliary field $b$ to lift the delta function,
and can get an equivalent Lagrangian
\begin{equation}   \label{LwithNLaf}
\begin{split}
&\mathcal{L}_{f}=\mathcal{L}_{\mathrm{YM}}+b(\partial^{\mu}A_{\mu})^{2}-\bar{c}Mc+BM\varphi-\bar{d}Md+\frac{1}{2}B^{2}.\\
&=\mathrm{Tr} \left(\!\!\!
\begin{array}{cc}
\mathcal{L}_{\mathrm{YM}} + b(\partial^{\mu}A_{\mu})^{2} - \bar{c}Mc  & 0 \\[1mm]
 0 & BM\varphi - \bar{d}Md + \frac{1}{2}B^{2}  \\
\end{array}
\!\!\! \right)=\mathrm{Tr}L  \, .
\end{split}
\end{equation}
Redefining the $s^{\prime}$ as:
\begin{equation}
\begin{split}
& s^{\prime} A_{\mu}^{a}=D_{\mu}^{a3}c, \;\;\; s^{\prime}b=0, \;\;\; s^{\prime}c=0, \;\;\; s^{\prime}\bar{c}=2b, \quad \\
& s^{\prime}\varphi=0, \qquad \;\; s^{\prime} d=0, \;\;\; s^{\prime}\bar{d}=0, \;\;\; s^{\prime} B=0, \;\;
\end{split}
\end{equation}
one can recognize easily that the newly defined $\mathcal{S}$ holds also the nilpotency
and the Lagrangian (in Eq.~(\ref{LwithNLaf})) abides definitely by well-defined nilpotent symmetry.

Besides the outcomes and properties we have discussed, we would like to remark another interesting issue, which has in fact encountered in last section.

We can reinterpret the core identity Eq.~(\ref{nid}) we proved in section IV in a different way. We denote $[A]_{s}=\{A^{\prime} =A^{U},\, U\in\mathcal{H}\}$, which is a suborbit of $[A]$.
Recall the definition of $N_{GC}(G(A))$, it is the intersection number between the orbit $[A]$ and the Landau hypersurface $R$ in configuration space $\mathfrak{A}$,
the identity can be considered as that the $[A]_{s}^{}$ crosses the Landau hypersurface once or less.
The validity of our approach requires that there should be at least one intersection for every orbit.
However, even if $[A]_{s}$ does not cross the Landau hypersurface straightforwardly,
the approach is still valid after employing the gauge transformation for $[A]_{s}$.
To see this distinctly we introduce an alternate definition for the Landau  hypersurface~\cite{semenov1986variational,zwanziger1982non}.
With defining the functional accordingly as
\begin{equation}
F_{A}(U)=\int\mathrm{d}^{4}xTr(A^{U}_{\mu}A^{U}_{\mu})
=\frac{1}{2}\int\mathrm{d}^{4}xA^{a,U}_{\mu}A^{a,U}_{\mu},
\end{equation}
one can easily verify that the Landau hypersurface $R$ coincides with the set of the relative extrema of $F_{A}(U)$ along $[A]$ (Furthermore, all the relative minima consist of Gribov region).
A straightforward conclusion is then that every gauge orbit $[A]$ intersects with $R$.
Consequently, for every configuration $A$, there exists $P(A)\in\mathcal{G}$ that satisfies $A^{P(A)}\in R$, the choice of $P(A)$ is arbitrary. Therefore, considering a replacement of identity Eq.~(\ref{nid})
\begin{equation}
1=\int_\mathcal{H}\mathcal{D}U\delta(G(A^{P(A)U}))|\Delta(G(A^{P(A)U}))|,
\end{equation}
which holds for every $A$, replacing the integral variables in generating functional as $A^{P(A)U}\rightarrow A$, and assuming there exist a specific choice of the functional $P(A)$ makes the Haar measure $\mathcal{D}U$ changing by a constant volume, one can recognize easily that the results in the previous section remain the same.

\section{Summary}

In this paper we investigate the modification of the Faddeev-Popov gauge fixing approach to non-Abelian gauge theory. After reviewing the Gribov problem~\cite{gribov1978quantization} arising in the Faddeev-Popov approach, we start our modification from the original idea of Faddeev and Popov by inserting a modified identity into the non-Abelian Yang-Mills full generating functional.
The identity is obtained by a group integral through the maximal Ablelian subgroup of the gauge group. Working in the simplest $SU(2)$ gauge theory, which can be easily generalised to $SU(N)$,
we prove that by this new approach, the problems of the Gribov ambiguity no longer exist.
By applying the method introduced in Ref.~\cite{ghiotti2005landau},
we get a local Lagrangian which is free from the Gribov ambiguity and meanwhile, the bare gluon propagator is well-defined, which has exactly the same expression as its counterpart in the standard Landau gauge of non-Abelian theory.
This fact reveals the efficiency of this new approach,
and indicates that extracting gauge parameter of the maximal Abelian subgroup is enough for eliminating the over-counting infinity from the original non-Abelian group.
Noticing that this brings about actually another issue, namely, how we deal with the remanent gauge configurations generated by the non-Abelian group generators
which cannot be described by the Maximum Abelian subgroup.
The clarification of this issue is in progress.
Here we just insist on the FP approach in which the inserted identity does not change the generating functional of the original QCD action, and thus, keep the remanent gauge configurations in the generating functional.
We also present the symmetry of the new Lagrangian, which can be viewed as an analogue to the BRST symmetry.

\medskip

The work was supported by the National Natural Science Foundation of China under Contracts No.~11435001 
and No.~11775041; the National Key Basic Research Program of China under Contracts No.~G2013CB834400
and No.~2015CB856900.

\end{document}